\input harvmac
\Title{\vbox{
\hbox{HUTP-98/A032}
\hbox{\tt hep-th/9804172}}}{Puzzles at Large N}
\bigskip
\centerline{Cumrun Vafa}
\bigskip
\centerline{Lyman Laboratory of Physics}
\centerline{Harvard University}
\centerline{Cambridge, MA 02138, USA}
\vskip .3in
We consider the AdS/CFT correspondence in the context
of 2d CFT and find that essentially
all ``single particle'' primary fields with higher spin, predicted from the CFT
side are missing from the Kaluza-Klein excitations of
the AdS supergravity.   The high mass extension of these missing states
gives rise to the macroscopic entropy of extremal 5d black holes.

\Date{April 1998}
\newsec{Introduction}
Recently there has been a renewed interest in studying the
large $N$
limit of quantum field theories.  In particular it was suggested
in \ref\jua{J. Maldacena,  hep-th/9711200.}\
following earlier work in connection with
black holes \ref\kelb{J. Maldacena and
 A. Strominger, Phys. Rev. {bf D56} (1997) 4975,
 hep-th/9702015;
   S.S. Gubser and  I.R. Klebanov,
 Phys. Lett. {\bf B413} (1997) 41,  hep-th/9708005 }\
that the fundamental strings in appropriate
backgrounds may provide the long sought after stringy interpretation
of large $N$ gauge theories in the regime of strong coupling
($g_sN \gg 1$).  This proposal has been sharpened in \ref\pol{
 S.S. Gubser, I.R. Klebanov and  A.M. Polyakov,
 hep-th/9802109.}\ref\witt{E. Witten, hep-th/9802150.}.
 In particular,  a precise relation between
Kaluza-Klein states of the supergravity in these
backgrounds and the spectrum of conformal operators at large $N$
was suggested.  In the case of ${\cal N}=4$ gauge theory in $d=4$
it was shown that the KK modes give rise
to all the operators of the form $Tr(X_{(i_1}...X_{i_n)})$
and its descendants \pol\witt \ref\fer{
 S. Ferrara, C. Fronsdal and  A. Zaffaroni, hep-th/9802203;
 Andrianopoli and S. Ferrara,
 hep-th/9803171.}\ref\oldsugra{H. Kim, L. Romans and
P. van Nieuwenhuizen, Phys. Rev. {\bf D32} (1985) 389;
M. Gunaydin and N. Marcus, Class. Quant. Grav. {\bf 2} (19985) L19;
 Class. Quant. Grav. {\bf 2} (1985) L11.} (see also the
related work \ref\oogh{ G.T. Horowitz and
 H. Ooguri,
 hep-th/9802116.}).  Of course, part of the conjecture is that
 the relation between large $N$ gauge theory and gravity modes
 is one to one.  In other words, for every conformal
operator one should find a state (in the fock space of)
KK excitations in supergravity.

Recently large $N$ limit of certain ${\cal N}=(4,4)$
gauge theories in 2 dimensions
was  considered from this point of view in \ref\mast{J. Maldacena and
A. Strominger, hep-th/9804085.}\ref\emi{E.J. Martinec, hep-th/9804111.}\
 and a large set of chiral primary operators, which can be viewed as
elements of one of the
$(c,c),(c,a),(a,c),(a,a)$ rings,
whose dimensions are protected from receiving quantum corrections
 were identified
with specific Kaluza-Klein excitations.  However as we shall see  a large
number of states
expected from the CFT side  are actually missing from the
Kaluza-Klein excitations.  The aim of this note is to elaborate
on these missing states.

\newsec{The Basic Puzzle}
In this section we state the basic puzzle in general terms.
The manifestation of this puzzle
 for the 2d is discussed in the subsequent section.

Consider a conformal theory in $d$ dimensions.  Let us consider
the Euclidean version.  The rotation  group is $SO(d)$, and the
conformal group is $SO(1,d+1)$.   The set of operators in this
theory will form representations of this group.  Let us consider
the primary operators of the conformal
group and ask what spins they carry under
$SO(d)$.  A priori one would expect that if we consider
sufficiently high dimension primary operators we should
be able to get sufficiently high spin representation of $SO(d)$.

We generally have  ``single particle'' operators which
correspond to a single particle state  in supergravity and multiparticle
operators which correspond to many particle states.
It is natural to  assume that single particle operators are represented in
the field theory by a single trace while multiparticle operators
correspond to operators with multiple traces (in the case of U(N)
super-Yang-Mills).
The AdS/CFT correspondence {\it predicts}
that as $N\rightarrow \infty$ (and $ (g_s N) \rightarrow \infty $)
 no matter which high dimensional primary operator we consider we
will never find a ``single particle''  operator with spin greater than 2!
This follows from the fact that when we expand the supergravity
modes and find the Kaluza
Klein states we are dealing with spins which go at most to spin
2. (Of course it is possible to have primary multiparticle operators
with spins greater than one).
This is a rather  counter-intuitive statement, at least for
2d CFT's.  In this case, if the central
charge $c$ is bigger than one then one gets infinitely
many primary states, and moreover,
their spins $L_0-{\overline L_0}$ are not bounded.
Sometimes what one can do, even if $c>1$
is to consider an enlarged symmetry algebra for which the primary
fields of the enlarged algebra are finite in number and all
have zero spin (for example the left-right symmetric
WZW models).  Note, however, that if we consider
a very large value of $c$, as is the analogue of the large $N$ theories,
for this to happen we need to have an infinitely big symmetry
algebra.  This is not the case for CFT's
 in the context of AdS/CFT correspondence
where the symmetry group is the superconformal
group, and which is independent of $N$. Thus
since we have a finite symmetry algebra we expect to get many primary
states with arbitrarily large spins.  This is the basic
puzzle.

\newsec{The 2d Example}

Here we consider type IIB strings and
the system of large number of 1-branes and
5-branes wrapped around $T^4$ or $K3$, recently analyzed
from the viewpoint of AdS/CFT correspondence in
\mast \emi .  We now discuss the missing
states in this case.
In the context of D1 and D5 branes
this system was already analyzed in \ref\vgas{C. Vafa,
 Nucl. Phys. {\bf B463} (1996) 415, hep-th/9511088;
Nucl. Phys. {\bf B463} (1996) 435, hep-th/9512078.}\
and applied to the question of microstates of 5 dimensional black holes
in \ref\strva{A. Strominger and  C. Vafa,
 Phys. Lett. {\bf B379} (1996),  hep-th/9601029. }.
 In particular the left- and right-moving
ground states of the left-over 1+1 CFT in the Ramond-Ramond sector
was analyzed in \vgas\ and was shown to correspond to BPS
states anticipated from string/string duality.  Moreover the
left-moving excited states and right-moving ground
states in the Ramond-Ramond sector, which have
non-zero spin $L_0-{\overline L}_0$ (and correspond
to non-zero momentum in spacetime) were used in \strva\
to account for microstates of certain extremal black holes
in 5 dimensions.  For that particular application only the
degeneracy of such states with large values of $L_0-{\overline L}_0$
was relevant, but as was argued \strva\ there is a
{\it lower bound}
on such values (at least in the $K3$ case) for {\it all}
values of $L_0-{\overline L}_0$ coming from the
consideration of elliptic genus of symmetric products\foot{
Here we mean the modified elliptic genus which on
the non-supersymmetric side measures the fermion number
by the insertion of ${\rm exp}(i\theta F_L)$ where
$F_L$ is the left-moving fermion number.}.
In particular
even for small values of $L_0-{\overline L}_0$ there is a non-zero
lower
bound for the number of such states (see  the analysis
of the elliptic genus of symmetric products of $K3$ in \ref\mdvv{
R. Dijkgraaf, G. Moore,  E. Verlinde and  H. Verlinde,
 Commun. Math. Phys. {\bf 185} (1997) 197, hep-th/9608096.}).
 Moreover this lower bound is independent of any deformations
of the theory, such as the  moduli of $K3$ or the coupling constant.
In the NS-NS sector these states get translated to states which
are chiral (or anti-chiral) primary on the right-mover side
but arbitrary state on the left-mover side.  The corresponding operators
form an infinite dimensional chiral algebra \ref\vaf{C. Vafa, {\it Superstring
Vacua}, Proc. of High Energy Physics and Cosmology
Summer School at ICTP, 1989.}.
 To get a feeling for
such states in the present example, consider the primary states
which are purely left-mover (i.e. tensored with the identity operator
on the right-mover side).  In particular consider $N$ symmetric
products of $K3$.  Let $T_i$ denote the left-moving energy momentum
tensor for the $i$-th $K3$. For the sigma model
on
$$(K3)^N/{S_N}$$
we can construct $N$ new holomorphic currents out of $T$, namely we
consider the permutation invariant operators
$$T^{l}=\sum_{i=1}^N T_i^l$$
for $l=1,...,N$.  It is easy to see for each $l$ (except
$l=1$) we can
form one primary field (using products of $T^{k_i}$ with
$\sum k_i =l$).
  In fact these currents lead to a kind of
$W$ algebra \ref\bak{I. Bakas, Comm. Math. Phys. {\bf 134} (1990) 487.}
(for a review of $W$ algebra see \ref\sez{E. Sezgin,
hep-th/9202086.} and references therein).
We can repeat this for all the ${\cal N} =4$
 superconformal generators and we obtain
some kind of ${\cal N}=4$ W-algebra.

We thus conclude that there are infinitely many primaries of the
${\cal N}=4$ superconformal algebras with arbitrary integer $L_0-{\overline
L}_0$ spin, in contradiction with anticipations based on gravity
which predicts only a small bounded set. In fact the states
that are missing from the AdS/CFT correspondence in this case
are huge and in the asymptotic
regime form the bulk of the microstates of the black
hole. In fact note that if the problem was only occuring for
extremely high values of spin, then one could have
perhaps matched it to some very high spin states
that are also expected in the AdS side, such as massive string states, etc.
 However we have
shown here that for any finite value of spin there are
such missing states.

Even though this example demonstrates that something is
missing in the AdS side to account for all the expected
CFT states, it is not clear from this example alone whether
or not this is connected with special features of conformal
theories in 2d.

\newsec{Discussion}

What we have shown here is that the proposed closed string theories on
AdS do not give a full account of all the states that we expect from the
2d CFT side.  That there should be some difficulty in getting
some states is perhaps to be expected.  After all, in the 2d example,
had we been able to identify all these higher spin states in a purely
gravitational setup we would have most probably been
able to account for the black
hole entropy directly in a semiclassical gravitational setup.

Our general discussion, as well as the concrete
example in the 2d case suggests that the organizing principle
for conformal fields at large $N$ should be a much bigger
algebra than simply the superconformal algebra.  We should perhaps
expect some
kind of infinite dimensional $W$ algebras to be present, such
as volume preserving diffeomorphisms in the corresponding
dimension.  However
if one is to get such a symmetry from an underlying closed string theory,
the corresponding gravity theory would have to involve something
analogous to $W_{\infty}$ gravity studied in the 2d case. In particular
the higher spin gravity states would couple to higher spin primary
fields, and the KK excitations of such a gravity
theory may have a chance of reproducing such states.  It is
an extremely interesting question to find the right `gravity' theory
which could account for the spectrum of all operators at large $N$.
It should be possible to make this more concrete in the context
of the large $N$ limit of 2d conformal field theories.  In fact in such
a case some connections between large $N$ and $W_\infty$
algebra and self-dual gravity in four dimensions has already been noted
\ref\ppar{Y.J. Park, Phys. Lett. {\bf B238} (1990)
287.}\ref\hull{C. Hull, Phys. Lett. {\bf B269} (1991) 257.}.
One would then expect that the large $N$ limit of
2d conformal theories should be
related to the $N=2$ string
\ref\oov{H. Ooguri and C. Vafa, Nucl. Phys. {\bf B361} (1991) 469.}\
which quantizes self-dual gravity in 2+2 dimensions.  It would
also be interesting
to find the analogous proposal in higher dimensions and also find
how it fits with the AdS/CFT correspondence.  Note that this
is giving a hint that a $d$ dimensional conformal theory may be
related to a theory in $d+2$ dimensions.  This is in accord with
the fact that $SO(d-1,1)$ Lorentz group is to be extended
to $SO(d,2)$ conformal group (i.e. it would be related to
including the extra dimension in which the AdS is embedded in).
This is somewhat reminiscent of F-theory.

One could also look for less radical solutions.
An interesting feature of the missing states is that the mismatch is not
so bad as it could be.
Consider all primary states with levels smaller than a given level $\Delta$.
Then the number of states is independent of $N$ (or $c$) for small
values of $\Delta$. This is not what we would expect for a generic
conformal field theory
(for example consider $c$ free bosons in 1+1 dimensions).
This is suggesting that probably the solution might be related to
understanding more properly the dynamics of the fields that might live
at the boundary of AdS. In fact higher spin fields were analyzed
in \ref\sez{ E. Bergshoeff, A. Salam, E. Sezgin and Y. Tanii,
Phys. Lett. {\bf B205} (1988) 237.} but these do not seem to be
the states we need because they can be viewed as two particle states.
Another possibility, in the 1+1 dimensional case, is that
 we have a Chern-Simons
theory  which might give some anyon statistics to the fields living in the
bulk of AdS. This might lead to new states coming from bound states of
the particle states that we already had.

Note that the question of extending the missing states to the higher
dimensional case, for example $N=4$ in four dimensions is a rather interesting
one (in fact the simplest guess for potentially missing states in
the form of ${\rm Tr} W_+^k$ where $W_+$ is the gaugino chiral field
suffers from the fact that they are not $N=1 $ primary\foot{We are
grateful to O. Aharony for pointing this out, and correcting
an error in the original version of this paper.}).  One way one
can imagine getting such states is by considering $N=4$ Yang-Mills
on $T^3$, and viewing the 4 dimensional theory from the $2d$ perspective
in which case we would expect to get similar operators
related to the elliptic genus.  It is conceivable
that they are related to `t Hooft flux operators in this context.

Despite the fact that we have found many states missing
on the AdS side in the 2 dimensional example, given the nice results that have
already been obtained from the conjectured AdS/CFT correspondence,
and the fact that there are no obvious states missing
in the higher dimensional cases, one
has the feeling that even if this proposal is not complete,
it is not too far from being complete.

I am grateful to J. Maldacena for many useful comments
and discussions.
I would also like to thank K. Intriligator, A. Johansen,
D. Kazhdan,  E. Martinec,
E. Sezgin, A. Strominger and E. Witten for valuable discussions.

This research was supported in part by NSF grant
PHY-92-18167.

\listrefs
\end